\documentclass[sigplan,10pt]{acmart}
\renewcommand\footnotetextcopyrightpermission[1]{}
\usepackage[ruled,linesnumbered, noend]{algorithm2e}
\usepackage{makecell}
\usepackage{booktabs}

\usepackage{amssymb}
\usepackage{bbding}

\usepackage{pifont}
\usepackage{graphicx} 
\usepackage{amsmath}
\usepackage{multirow}
\usepackage{subfigure}
\usepackage{makecell}
\usepackage[ruled,linesnumbered, noend]{algorithm2e}
\usepackage{enumitem}
\usepackage{xcolor}
\usepackage{lipsum}
\usepackage{comment}
\usepackage{hyperref}
\usepackage{listings}

\AtBeginDocument{%
  \providecommand\BibTeX{{%
    \normalfont B\kern-0.5em{\scshape i\kern-0.25em b}\kern-0.8em\TeX}}}

\copyrightyear{2026}
\acmYear{2026}
\acmConference[arXiv preprint]{}
\acmDOI{}
\acmISBN{}

\title{EngineAgent: Agentizing On-Device LLM Inference Engines with Inference Stack Hierarchy Awareness and Hardware Abilities}

\author{Zhengxiang Huang}
\orcid{0009-0000-1536-1616}
\affiliation{%
  \institution{Shanghai Jiao Tong University}
  \city{Shanghai}
  \country{China}
}

\author{Yujie Sun}
\orcid{0009-0000-3314-5159}
\affiliation{%
  \institution{Shanghai Jiao Tong University}
  \city{Shanghai}
  \country{China}
}

\author{Shengheng Chen}
\orcid{0009-0003-7249-947X}
\affiliation{%
  \institution{Shanghai Jiao Tong University}
  \city{Shanghai}
  \country{China}
}

\author{Zhaode Wang}
\orcid{0009-0009-0498-7491}
\affiliation{%
  \institution{Alibaba Group}
  \city{Hangzhou}
  \country{China}
}

\author{Chaoyue Niu}
\authornote{Chaoyue Niu is the corresponding author (rvince@sjtu.edu.cn).}
\orcid{0000-0002-1650-4233}
\affiliation{%
  \institution{Shanghai Jiao Tong University}
  \city{Shanghai}
  \country{China}
}

\author{Zeyu Zhao}
\orcid{0009-0007-5768-166X}
\affiliation{%
  \institution{Shanghai Jiao Tong University}
  \city{Shanghai}
  \country{China}
}

\author{Chengfei Lv}
\orcid{0009-0004-6618-521X}
\affiliation{
  \institution{Alibaba Group}
  \city{Hangzhou}
  \state{Zhejiang}
  \country{China}
}

\author{Fan Wu}
\orcid{0000-0003-0965-9058}
\affiliation{%
  \institution{Shanghai Jiao Tong University}
  \city{Shanghai}
  \country{China}
}

\author{Guihai Chen}
\orcid{0000-0002-6934-1685}
\affiliation{%
  \institution{Shanghai Jiao Tong University}
  \city{Shanghai}
  \country{China}
}

\newcommand{\systemname}{EngineAgent\xspace}

\begin{document}

\begin{abstract}
Rapidly evolving LLM architectures and heterogeneous mobile hardware make supporting new models in on-device inference engines a recurring and costly system challenge. 
Each model may require coordinated changes across conversion, graph construction, operators, and hardware backends. Existing coding agents often miss these cross-layer dependencies and struggle to diagnose failures across host--device execution paths.
We present \emph{EngineAgent}, an agentic system for automating on-device LLM model enablement. Its key insight is to separate reusable engine and hardware knowledge from model-specific implementation. 
Before enablement, \emph{EngineAgent} constructs a hierarchy-aware engine index that captures engine abstraction layers and cross-layer dependencies, and pre-validates backend-specific build, execution, and debugging workflows using already supported models.
For each new model, a builder uses this knowledge to coordinate cross-layer changes, while an auditor checks implementation completeness and drives iterative debugging through device-side validation.
Across 12 tasks spanning three mobile LLMs and four engine--backend pairs on CPUs, GPUs, and NPUs, \emph{EngineAgent} achieves 75\% end-to-end success with either of two underlying LLMs, versus 33.3\% for the strongest coding-agent baseline. Success requires compilation, on-device execution, and output-logit agreement with reference implementations. \emph{EngineAgent} also reduces average per-task time and model-API cost relative to baselines using the same underlying LLMs.
\end{abstract}



\maketitle

\section{Introduction}
\label{sec:intro}
On-device LLM inference engines are critical middleware positioned between two distinct ecosystems: \emph{rapidly evolving LLM model architectures} above and \emph{a fragmented landscape of heterogeneous mobile hardware} below. 
Engine developers must continuously extend their engines to support previously unsupported model architectures and capabilities while ensuring their correct execution across underlying diverse mobile CPUs, GPUs, and NPUs.

On the model side, device-oriented lightweight LLM families evolve rapidly, with successive generations often released only months or even weeks apart. For instance, Gemma~3 to Gemma~3n only took about 4 months~\cite{Gemma3,Gemma3n}, while MiniCPM-V~4.0 to 4.5 took only 24 days~\cite{MiniCPM-V-4.5}. More importantly, successive generations frequently introduce new modules and designs. Qwen3.5~\cite{Qwen3.5} introduces GatedDeltaNet linear attention~\cite{GatedDeltaNet} interleaved with conventional attention. Gemma4~\cite{Gemma4} adds hybrid sliding-window/global attention, proportional RoPE (p-RoPE), unified keys and values, shared KV cache,  and Per-Layer Embeddings (PLE). MiniCPM-V~4.6 further introduces a window-attention vit\_merger, intra-ViT early compression, and mixed $4\times/16\times$ visual-token compression~\cite{MiniCPM-V-4.6}. Consequently, supporting one generation rarely guarantees support for its successors, forcing inference engines to repeatedly incorporate newly introduced model structures and modules.

On the hardware side, mobile deployment is fragmented across heterogeneous execution backends. ARM CPUs, mobile GPUs~\cite{Mali}, Qualcomm Hexagon NPUs~\cite{hexagon-npu, hexagon-htp}, MediaTek APUs~\cite{NeuroPilot}, and Apple Neural Engine~\cite{CoreML} differ substantially in programming languages and interfaces, tensor layouts, compilation workflows, and runtime semantics. 
Therefore, supporting the same new LLM feature requires substantially different implementation paths across hardware backends. CPU backends may implement new operators directly in C/C++ or assembly; Vulkan GPUs may require dedicated compute shaders, SPIR-V compilation, and backend-specific dispatch logic; NPUs typically require graph rewriting, accelerator-specific lowering, and vendor-specific compilation.
Consequently, each newly introduced model feature need to be independently adapted and validated for multiple backends through entirely different engineering paths, multiplying the implementation efforts required for enablement.

\begin{figure*}[t]
    \centering
    \includegraphics[width=0.9\linewidth]{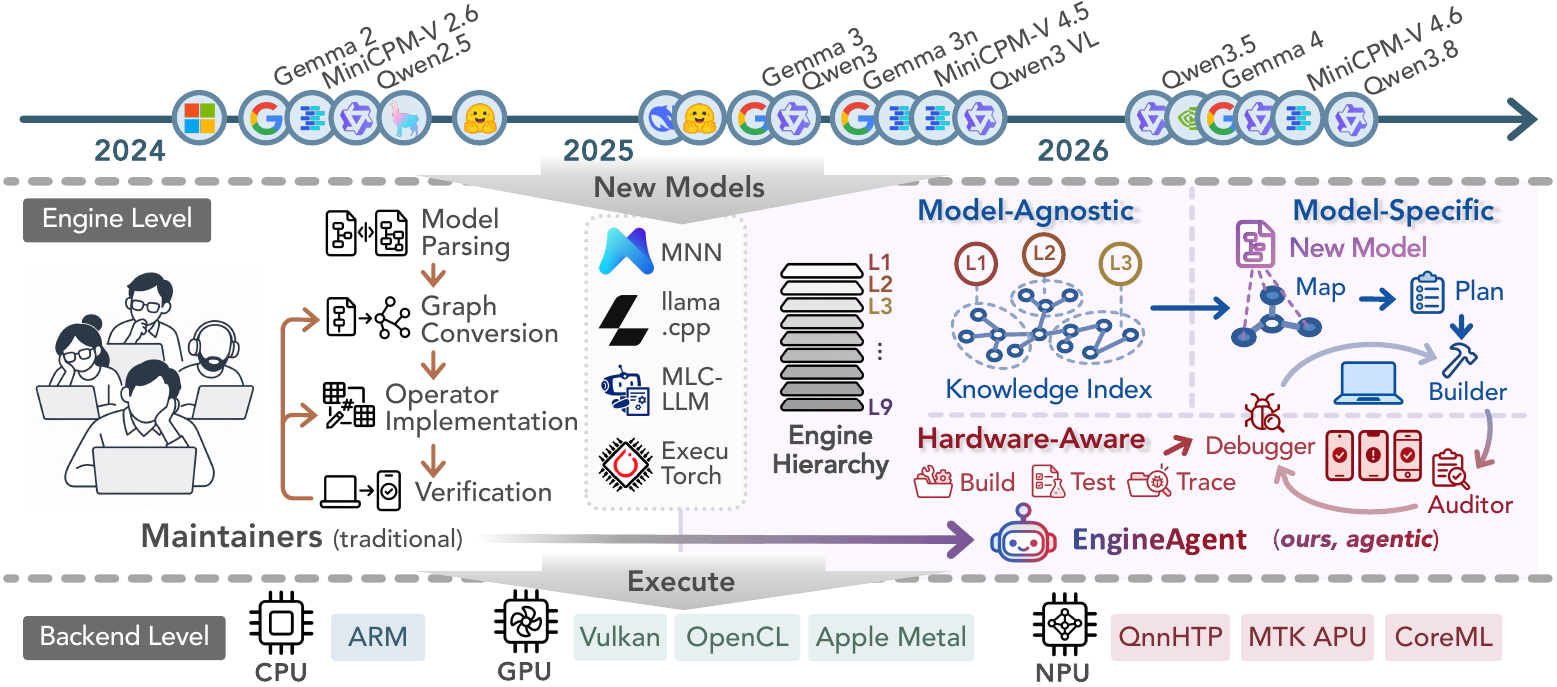}
    \caption{
        Position and workflow of \systemname: agentically mapping rapidly evolving LLM architectures onto heterogeneous mobile hardware for inference engines.
    }
    \label{fig:motivating-cases}
    \vspace{-2mm}
\end{figure*}

Traditionally, enabling a new model remains a largely handcrafted, expert-driven process, requiring expertise spanning both LLM model architecture and thorough understanding of the inference engine stack. 
Maintainers must analyze the new model semantics, map new model features onto engine modules, modify export and runtime execution paths, adapt operators or kernels, and validate execution on target backends. 
For example, enabling Qwen3.5-2B in MNN~\cite{Walle} touched 42 files and over 4.2K lines across export and execution runtime, shape inference, and both CPU and GPU backends, whereas ExecuTorch~\cite{ExecuTorch} introduced changes to 20 files and over 1K lines for text-only CPU support. 
Such broad and heavy modifications impose substantial development and validation burdens on engine maintainers, which are also reflected in both \emph{incomplete backend and modality coverage} and \emph{enablement delay}. 
As summarized in Table~\ref{tab:model-enablement-delay}, 2 out of the 4 on-device engines do not support Gemma4-E2B-it, while 3 out of the 4 do not support MiniCPM-V~4.6. Even when a model is supported, coverage is often restricted to only a subset of backends or modalities, such as CPU-only or text-only execution. 
Moreover, initial on-device enablement can take up to 30 days after model release.

The substantial engineering burden naturally motivates us to \emph{agentize} model enablement. Before a new model is released, agents should acquire reusable knowledge of the engine's semantics and establish backend-specific debugging and verification workflows using already supported models. After release, agents should analyze the new model's architecture, use this knowledge to identify and implement the required engine modifications, and iteratively debug and verify the implementation on real devices.


\begin{table}[t]
\centering
\caption{Model enablement\protect\footnotemark{} backend coverage and delay across on-device LLM inference engines~\cite{ExecuTorch, Walle, llama.cpp, MLC-LLM}.}
\label{tab:model-enablement-delay}
\small
\resizebox{\linewidth}{!}{%
\begin{tabular}{lllll}
\toprule
\textbf{Model} & \textbf{Release} & \textbf{Engine} & \textbf{First Enablement} & \textbf{Backend} \\
\midrule
\multirow{4}{*}{Qwen3.5-2B}
& \multirow{4}{*}{2026-03-02}  & ExecuTorch & 04-01 / 30 days & CPU (text-only) \\
&                                      & MNN        & 03-02 / 0 day & CPU / GPU / NPU \\
&                                      & llama.cpp  & 03-02 / 0 day & CPU\\
&                                      & MLC-LLM    & 03-08 / 6 days & GPU\\
\midrule
\multirow{4}{*}{Gemma4-E2B-it}
& \multirow{4}{*}{2026-04-02}  &  ExecuTorch & N.A. & unsupported \\
&                                      & MNN        & 04-09 / 7 days & CPU \\
&                                      & llama.cpp  & 04-02 / 0 day & CPU \\
&                                      & MLC-LLM    & N.A. & unsupported \\
\midrule
\multirow{4}{*}{MiniCPM-V 4.6}
& \multirow{4}{*}{2026-05-11} & ExecuTorch & N.A. & unsupported \\
&                                      & MNN        &  N.A. & unsupported \\
&                                      & llama.cpp  & 05-11 / 0 day & CPU \\
&                                      & MLC-LLM    &  N.A. & unsupported \\
\bottomrule
\end{tabular}%
}
\end{table}

\footnotetext{The enablement date is defined as the date first verified releases or pull requests become public available.}

\paragraph{Challenges of Agentization.}
However, both localizing modifications within an engine and debugging and verifying them on real devices are fundamentally challenging.
\emph{\ding{172} Challenges of mapping new model features across hierarchical engine abstractions.}
On-device LLM engines such as llama.cpp~\cite{llama.cpp}, MNN~\cite{Walle}, ExecuTorch~\cite{ExecuTorch}, and MLC-LLM~\cite{MLC-LLM}, share a complex hierarchical organization that can be abstracted into multiple levels with distinct representations and manipulation logic.
At the input level, natural-language or visual inputs are transformed through tokenization or patchification; at the graph level, tensor computation is constructed, converted, fused, and scheduled as computation graphs; and at the operator level, computations are realized through backend-specific kernels and runtime dispatch.
During model enablement, a new feature may require coordinated modifications across these levels, yet their respective roles are not apparent from code structure alone.
Without understanding where a component lies in the engine hierarchy and what role that level plays in model execution, an agent may identify and modify locally relevant code while overlooking modifications required at other levels.
Therefore, reliable modification localization requires reasoning about the engine hierarchy rather than treating the codebase as a flat collection of files and symbols. \emph{\ding{173} Challenges of cross-device debugging and verification.} Unlike conventional host-only software, where execution and debugging typically occur within the same environment, on-device model enablement follows a host--device, cross-language, build--debug pipeline spanning host-side preparation and device-side deployment and execution. As a result, a failure observed on device may originate from an earlier host-side stage rather than from the failing runtime itself. Moreover, mobile CPU, GPU, and NPU backends follow distinct compilation and runtime workflows, each requiring different debugging methods. Consequently, tracing failures back to their root causes across devices and backends is inherently challenging. 



\paragraph{Why Existing Coding Agents Fail.}
Existing repository-level coding agents struggle with both challenges.
\emph{\ding{172} Code structure centric knowledge rather than engine hierarchy knowledge.}
Existing agents~\cite{LocAgent, CodeAgent, CodePlan, OpenHands} primarily reason over files, functions, symbols, and AST-level dependencies without explicitly modeling how these entities related to engine abstraction hierarchy. 
Consequently, their reasoning and modifications are typically constrained to local changes of one or two modules, missing necessary modifications across distant abstraction layers, e.g., adding an operator definition without implementing its hardware kernel.
\emph{\ding{173} Host-centric verification and debugging rather than cross-language, host--device, backend-specific workflows.}
Existing agents largely rely on language-specific, host-local static analysis, unit tests, and debug tracing~\cite{AutoCodeRover, AgentDebug}. 
They lack comparable capabilities for orchestrating and reasoning over host--device debug workflows, causing device deployments to fail, device-side failures to remain unattributed or misattributed, and debugging to fall back to repeated build--test attempts without precise localization. Even Codex~\cite{Codex} powered by GPT-5.6-Sol exhibits both missing implementations and hardware-verification failures (Figure~\ref{fig:failure-analysis}); for example, when enabling MiniCPM-V~4.6 on MNN--CPU, it stopped after an incomplete build and skipped device validation (Table~\ref{tab:agent-failures}).


\paragraph{Our System.}
We propose \systemname, an agentic system for reliable on-device LLM engine model enablement that automates the workflow while addressing the two challenges above.
\systemname consists of three key designs: a \emph{model-agnostic engine hierarchy knowledge index}, \emph{hardware-aware debug skills}, and a \emph{model-specific multi-agent implementation--audit main loop}.
To address the lack of engine-hierarchy understanding, \systemname constructs a hierarchy-aware knowledge index (\S \ref{subsec:hierarchy-aware-design}) that maps code structures onto engine abstraction layers and cross-layer dependencies, enabling agents to identify complete modification scopes beyond local code relations.
To address unreliable hardware-grounded debugging, \systemname constructs reusable build, test, and debugging skills (\S \ref{subsec:hardware-aware-design}) by validating backend-specific workflows with already supported models, providing stable verification procedures for heterogeneous hardware.
These two designed artifacts are pre-constructed ahead of time because they are reusable and model-agnostic. Moreover, they are designed to be updatable upon new engine commits.
When a new model arrives, it enters the model-specific multi-agent enablement main loop (\S \ref{subsec:main-loop}), where newly introduced model features are mapped onto the pre-constructed hierarchy knowledge index and hardware-aware debug skills.
The builder then plans and implements layer-wise modifications guided by the hierarchy index, and submits each completed change to the red-team auditor.
If validation fails, failures are routed to the builder with hardware-aware debugging support for iterative refinement.
The auditor accepts the implementation only after all layer-wise checks and end-to-end device validation pass.

Together, \systemname{} supports reliable model enablement across heterogeneous inference engines and hardware backends.
We evaluate \systemname on 12 model-enablement tasks spanning 3 mobile LLMs and 4 engine--backend pairs that cover CPU, GPU, and NPU execution: MNN--CPU, llama.cpp--Vulkan, MLC-LLM--OpenCL, and ExecuTorch--QNN. \systemname achieves an end-to-end success rate of \textbf{75\%}, compared with \textbf{33.3\%} for the strongest coding-agent baseline.

This work makes the following contributions:
\begin{itemize}
\item We identify \emph{model enablement} as a distinct cross-layer software-engineering problem for on-device LLM inference engines, and characterize its two key challenges: \emph{hierarchical system understanding} and \emph{hardware-aware debugging and verification} across heterogeneous devices and backends.

\item We design \systemname, an agentic system that addresses these challenges through three key components: a \emph{model-agnostic hierarchy-aware engine knowledge index} that captures abstraction layers and cross-layer dependencies, \emph{hardware-aware debug skills} that provide reusable and validated build--debug workflows, and a \emph{model-specific main loop} that maps new model features onto these pre-constructed artifacts for layer-wise implementation, completeness auditing, device-side validation, and iterative refinement.

\item We evaluate \systemname on 12 model-enablement tasks across three mobile LLMs and four engine--backend pairs, demonstrating higher effectiveness and efficiency than existing coding agents across benchmark cases.
\end{itemize}

\section{Background}
\label{sec:background}
\subsection{On-Device LLM Inference Engines} \label{subsec:mobile-engines}
On-device LLM inference engines provide the unified software infrastructure for executing diverse LLMs on mobile devices equipped with heterogeneous processors such as CPUs, GPUs, and NPUs.
Representative engines, including MNN~\cite{Walle, MNN-LLM}, ExecuTorch~\cite{ExecuTorch}, llama.cpp~\cite{llama.cpp}, and MLC-LLM~\cite{MLC-LLM}, share a common two-stage architecture: an \emph{offline, on-host model export stage}, realized by the engine's converter and compiler, and an \emph{online, on-device execution stage}, powered by the engine's runtime.
Orthogonally, both stages are organized into the same three levels of abstraction: the \emph{input level} (chat templates and tokenizers), the \emph{graph level} (the computation graph and its intermediate representation, IR), and the \emph{operator level} (backend kernels and their hardware mappings).

\begin{figure}
    \centering
    \includegraphics[width=0.8\linewidth]{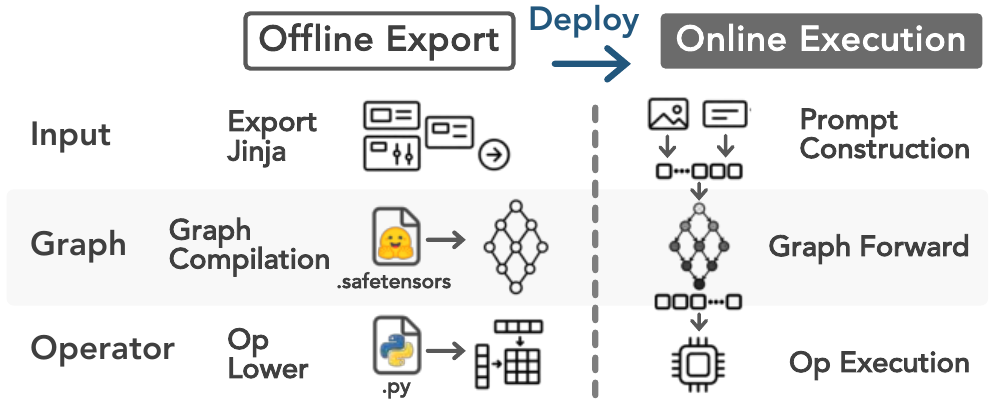}
    \caption{On-device LLM engines comprise export and execution stages, which share three abstraction layers.}
    \label{fig:background}
\end{figure}

As is illustrated in Figure \ref{fig:background}, the \emph{offline export stage} traverses these three levels top-down. At the input level, it exports the chat template and the (multi-modal) tokenizer and packs them into the model artifact. 
At the graph level, it maps PyTorch layers and modules onto engine-native IR nodes, applying transformations such as quantization, weight packing, layout reordering, and operator fusion. 
At the operator level, it lowers and links each IR node to a hardware-specific kernel from the engine's backend operator library.
Notably, certain engine--hardware pairs defer preparation to runtime via online just-in-time (JIT) compilation, such as the OpenCL backend for mobile GPUs in MNN but the JIT workflow mirrors its offline counterpart level for level.
The \emph{online execution stage} traverses the same levels at inference time. 
Engine runtime loads the prepared artifact, applies input-level pre-processing and tokenization to the user input, then walks the graph node by node, dispatching each node to its operator implementation. 

In summary, although these engines differ in their IR designs and supported backends, they converge on the same two-stage,
three-level workflow. 
This shared abstraction is what allows our analysis to generalize across engines rather than being tied to any single instance.

\subsection{Model Enablement from Maintainers' View} 
\label{subsec:model-enablement} 
\begin{figure}[t]
    \centering
    \subfigure[Example: mapping Qwen3.5-2B~\cite{Qwen3.5} features to MNN~\cite{Walle} abstraction layers.]{
        \includegraphics[width=0.85\linewidth]{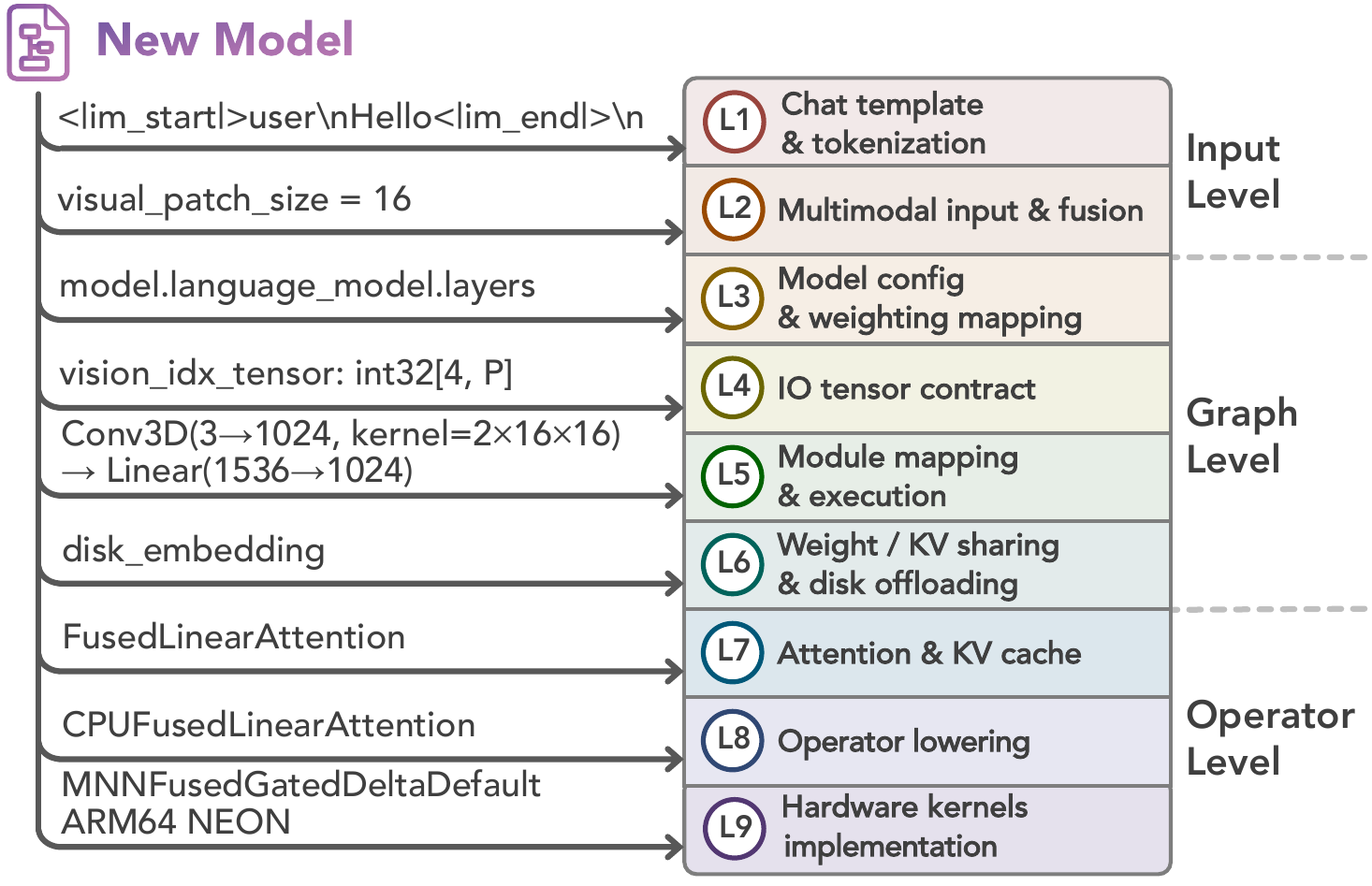}
        \label{fig:layer-mapping}
    }

    \subfigure[Host-device build-verify-debug loop.]{
        \includegraphics[width=0.75\linewidth]{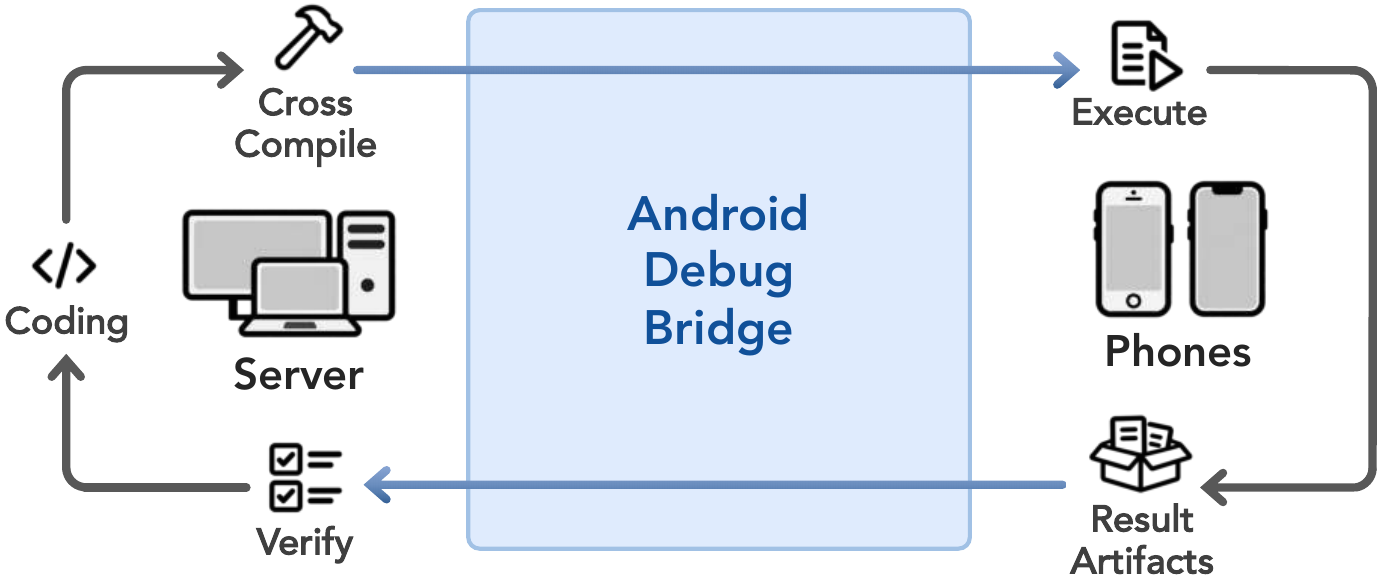}
        \label{fig:hardware-loop}
    }
    \caption{Model enablement workloads: mapping new model features to engine abstraction layers and then perform host-device build-verify-debug loop.}
    \label{fig:model-enablement}
\end{figure}

For engine maintainers, supporting a new LLM in a mobile inference engine is a complicated engineering effort, involving two primary procedure: understanding the engine and mapping new model features to modifications across engine modules and abstraction layers, and perform the host-device coordinated verify-debug loop to debug. Figure \ref{fig:model-enablement} exemplifies these 2 major procedure.

Maintainers first analyze newly introduced model features and determine how they map onto the engine's abstraction levels and modules. 
As illustrated by the example of mapping Qwen3.5-2B~\cite{Qwen3.5} to MNN~\cite{Walle} in Figure~\ref{fig:layer-mapping}, changes such as new chat-template formats or multimodal settings (e.g., "\texttt{<|im\_start|>}" and input visual patch size) are mapped to the input level; new model structures and tensor contracts such as the vision index tensor shape and data type are mapped to the graph level;  newly introduced operations such as \texttt{FusedLinearAttention}~\cite{GatedDeltaNet} must be lowered to operator level backend specific implementations, e.g., \texttt{CPUFusedLinearAttention} and its ARM64 NEON implementations. 
After the cognitive step of these mappings from model features to abstraction layers, the maintainers implements modifications to files and modules in each layer. 

Maintainers' implementation loop is an iterative host--device verify-debug loop. 
As shown in Figure~\ref{fig:hardware-loop}, they first modify on host across Python, C/C++, assembly, or backend specific code, then cross-compile it using the corresponding toolchains, such as the Android NDK, Vulkan toolchain, or QNN SDK. 
The resulting binaries and model artifacts are deployed to the target phone through Android Debug Bridge (ADB) for execution, after which logs, outputs, and other runtime artifacts are collected back on the host for verification. 
If any stage fails, maintainers trace the observed error back to the responsible implementation, revise the code, and repeat the coding--build--deploy--execute--verify cycle until all required checks pass.

\subsection{Agent for Systems}
\label{subsec:agent-for-systems}
To alleviate the substantial human effort in model enablement, agentizing the workflow is a natural direction.
Existing works have developed complementary capabilities for software engineering: general-purpose end-to-end coding agents directly solve development tasks through iterative code modification and execution, while code-analysis tools and agents enhance end-to-end coding agents with repository representations, modification localization, and edit plans, and debugging agents augment them with failures diagnosis and generate corrective feedback.
However, these capabilities do not directly resolve the challenges of inference-engine model enablement.

\paragraph{Limitations of General Purpose End-to-End Coding Agents.}
CodeAgent~\cite{CodeAgent}, SWE-Agent~\cite{SWE-Agent}, and OpenHands \cite{OpenHands} provide end-to-end workflows integrating repository navigation, code modification, execution, and iterative feedback.
Recent systems efforts such as SYSSPEC~\cite{SYSSPEC} and ECO \cite{ECO} further incorporate domain-specific abstractions and production constraints into LLM-driven system automation.
However, these approaches still do not target inference-engine model enablement as an end-to-end task: they neither expose engine hierarchies as first-class knowledge nor provide dedicated host--device verification and debugging procedures across heterogeneous hardware backends.

\paragraph{Limitations of Prior Code Analysis Tools and Agents.}
CodeGraph \cite{CodeGraph} and LocAgent~\cite{LocAgent} improve repository representation and localization, CodePlan~\cite{CodePlan} plans coordinated repository-wide edits, and AutoCodeRover~\cite{AutoCodeRover} combines program-structure-aware search with fault localization.
However, these methods primarily organize knowledge around code-level structures such as files, symbols, functions, and AST or dependency relations.
They do not explicitly model how these entities map onto inference-engine abstraction layers, making cross-layer dependencies, wrong-layer localization, and structurally distant but semantically coupled modifications difficult to reason about.

\paragraph{Limitations of Prior Debug Agents.}
AgenTracer~\cite{AgenTracer} attributes failures to responsible agents and trajectory steps, while AgentDebug~\cite{AgentDebug} identifies root causes and provides corrective feedback for recovery.
Such approaches improve failure analysis, but remain largely centered on agent trajectories or host-side program feedback.
Model enablement instead requires cross-language, cross-compilation debugging across host preparation, device deployment, and heterogeneous XPU execution, where failures must be traced from device-side observations back to the responsible host-side implementation.

\section{Case Studies and Key Insights}
\label{sec:insights}

\subsection{Characterizing Agent Failures}
\label{subsec:characterization}

\paragraph{Motivating Case Studies.}
To understand the limitations of existing coding agents for on-device LLM model enablement, we conduct motivating case studies on three representative engine--backend pairs: MNN--CPU, llama.cpp--Vulkan, and ExecuTorch--QNN, with MiniCPM-V~4.6 as the target model.
We evaluate representative repository-level coding agents, including SWE-Agent~\cite{SWE-Agent}, OpenHands~\cite{OpenHands}, CodePlan~\cite{CodePlan}, OpenCode~\cite{OpenCode}, Codex, and CodeGraph~\cite{CodeGraph} + OpenCode + AgenTracer~\cite{AgenTracer}.
The evaluated agents use their corresponding backbone models: DeepSeek-V4-Flash for open-source agent frameworks and GPT-5.6-Sol for Codex~\cite{Codex}.

Across these cases, existing agents demonstrate promising capability in repository exploration and local code modification, but rarely complete the entire model enablement workflow from model adaptation to validated on-device execution.
Across 18 test cases (6 agents $\times$ 3 representative engine--backend pairs), only 1 attempt successfully completes end-to-end model enablement. 
The remaining 17 attempts fail at different stages, including cross-layer implementation, backend integration, and device-grounded verification.
Notably, the only successful case, OpenCode on MNN--CPU, supports the CPU backend, which has a simpler lowering and compilation stack than GPU/NPU. 
These observations motivate a systematic categorization of agent failures.

\begin{table}[t]
\centering
\caption{Representative failure cases of existing coding agents during on-device LLM model enablement.}
\label{tab:agent-failures}
\small
\setlength{\tabcolsep}{3pt}
\renewcommand{\arraystretch}{1.15}

\resizebox{\linewidth}{!}{
\begin{tabular}{clccl}
\toprule
\makecell{\textbf{Engine}\\\textbf{Backend}} &
\textbf{Agent} &
\makecell{\textbf{Failed}\\\textbf{Stage}} &
\makecell{\textbf{Failure}\\\textbf{Category}} &
\textbf{Symptom} \\
\midrule

\multirow{5}{*}{\makecell{MNN\\CPU}}
& CodePlan
& Planning
& \makecell{Missing\\implementation}
& \makecell[l]{Localized config changes\\but missed dependent vision\\and runtime implementations.}
\\

& SWE-Agent
& Implementation
& \makecell{Missing\\implementation}
& \makecell[l]{Modified model components\\but missed export--runtime\\propagation.}
\\

& Codex
& Verification
& \makecell{Premature\\acceptance}
& \makecell[l]{Stopped after incomplete build\\and skipped device validation.}
\\

\midrule

\multirow{4}{*}{\makecell{llama.cpp\\Vulkan}} & OpenHands
& \makecell{Backend\\integration}
& \makecell{Backend\\ failure}
& \makecell[l]{Failed to resolve Vulkan\\compilation and deployment.}\\

& OpenCode
& \makecell{Backend\\integration}
& \makecell{Backend\\failure}
& \makecell[l]{Host adaptation completed\\but Vulkan execution path\\remained incomplete.}
\\

\midrule

\multirow{3}{*}{\makecell{ExecuTorch\\QNN}} & CodePlan
& Planning
& \makecell{Missing\\implementation}
& \makecell[l]{Followed local dependencies\\but missed QNN export.}
\\

& OpenHands
& \makecell{Backend\\integration}
& \makecell{Backend\\ failure}
& \makecell[l]{Failed at QNN lowering\\and runtime constraints.}
\\

\bottomrule
\end{tabular}
}
\end{table}

\paragraph{Agent Failures Categorization.}
We summarize representative failure cases of existing coding agents in Table~\ref{tab:agent-failures}.
Although the failures occur at different stages and across different engine--backend pairs, they exhibit two recurring patterns.
First, agents frequently identify and modify locally relevant components but fail to recover the complete implementation scope required by a new model feature.
For example, CodePlan on MNN--CPU modified model-related configuration paths but missed the dependent vision and runtime implementations, while SWE-Agent modified model components without propagating changes across export and execution paths.
Second, agents often achieve intermediate host-side progress but fail to complete hardware-grounded validation.
For example, OpenCode and OpenHands on llama.cpp--Vulkan reached host-side adaptation but failed to establish the complete Vulkan execution workflow, while Codex prematurely stopped after an incomplete build state without reaching device validation.
These failure patterns indicate that reliable model enablement requires both engine-level understanding across abstraction layers and hardware-grounded verification capabilities.

\subsection{Key Insights}
\label{subsec:insights}

The observed failures reveal two fundamental requirements for reliable agentic model enablement: global understanding of inference-engine abstractions and stable hardware-grounded verification workflows.

\paragraph{Insight 1: Model Enablement Agent Requires Engine Hierarchy Aware Understanding.}
Missing implementations and premature exits often stem from agents lacking a global view of the inference engine organization.
Without understanding the responsibilities and dependencies of different abstraction layers, agents may treat locally discovered code paths as the complete implementation scope and overlook required modifications elsewhere.
Therefore, agents require explicit hierarchy-aware engine knowledge to map model features to affected layers, recover complete modification scopes, and guide cross-layer implementation.

\paragraph{Insight 2: Model Enablement Agent Requires Hardware Aware Debug Ability.}
Hardware-aware debugging is challenging not only because device execution is heterogeneous, but also because the verification path itself may be unstable when enabling previously unsupported models.
Agents that directly debug unsupported models must simultaneously resolve model adaptation issues and uncertain build, deployment, and validation workflows, making failures difficult to attribute.
Therefore, reliable enablement requires pre-constructed and validated build, test, and debugging workflows from already supported models, which provide stable hardware-grounded feedback for subsequent model adaptation.

\section{System Design}
\label{sec:design}

\subsection{System Overview}\label{subsec:overview}

Guided by the two key insights, \systemname{} augments general-purpose coding agents with two capabilities: \emph{engine-hierarchy-aware understanding} for complete cross-layer implementation and \emph{hardware-aware debugging and verification} for reliable on-device validation. These capabilities support three complementary agent roles: a \emph{builder} that implements model support, a \emph{debugger} that diagnoses failures, and an independent \emph{auditor} that challenges implementation completeness and correctness.

Following Insight 1, \systemname{} equips agents with a model-agnostic \emph{engine hierarchy knowledge index} (\S\ref{subsec:hierarchy-aware-design}).
The index connects engine abstractions to concrete code entities and exposes dependencies across export, runtime, and backend layers. It guides the builder in mapping new model features to the required modification scope and provides the auditor with structural context for identifying omitted implementations. Following Insight 2, \systemname{} equips agents with reusable \emph{hardware-aware verify-debug skills} (\S\ref{subsec:hardware-aware-design}). Before a new model arrives, construction agents establish and validate backend-specific build, deployment, testing, and debugging workflows using already supported models. These skills provide the builder with self-validation procedures, the debugger with diagnostic guidance, and the auditor with executable checks, reducing uncertainty in the verification path during new-model adaptation.

\begin{figure}[t]
\centering
\includegraphics[width=\linewidth]{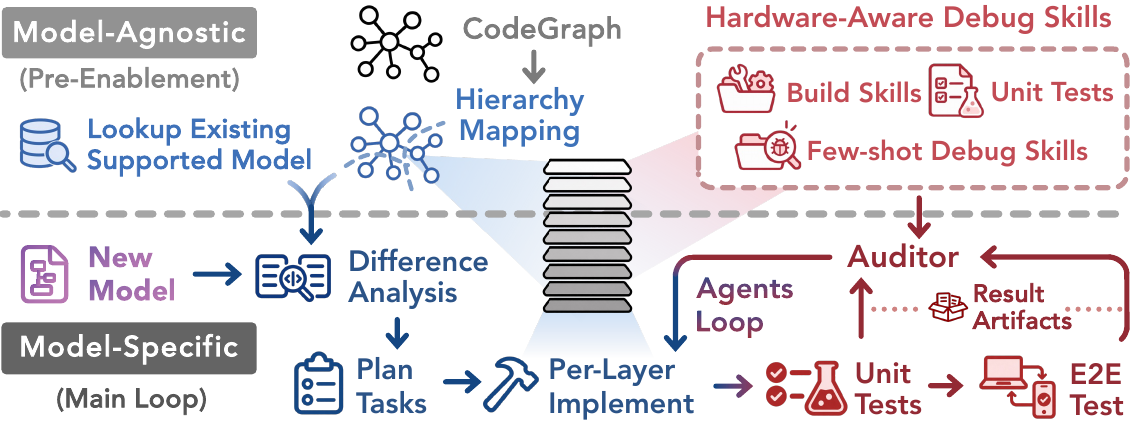}
\caption{Two-stage workflow of \systemname, from engine-specific knowledge construction to model-dependent enablement and verification.}
\label{fig:pipeline}
\end{figure}

As illustrated in Figure~\ref{fig:pipeline}, both capabilities are prepared before model release and reused in the model-specific multi-agent main loop (\S\ref{subsec:main-loop}). When a new model arrives, the builder compares its architecture with supported models, maps the differences onto the hierarchy index, and implements layer-wise modifications. The auditor checks each completed layer, while the debugger uses the validated host--device workflows to diagnose failures and guide revisions. The loop continues until the implementation passes both layer-wise checks and end-to-end device validation.

\subsection{Hierarchy-Aware Engine Understanding}

\begin{figure}[h]
    \centering
    \includegraphics[width=\linewidth]{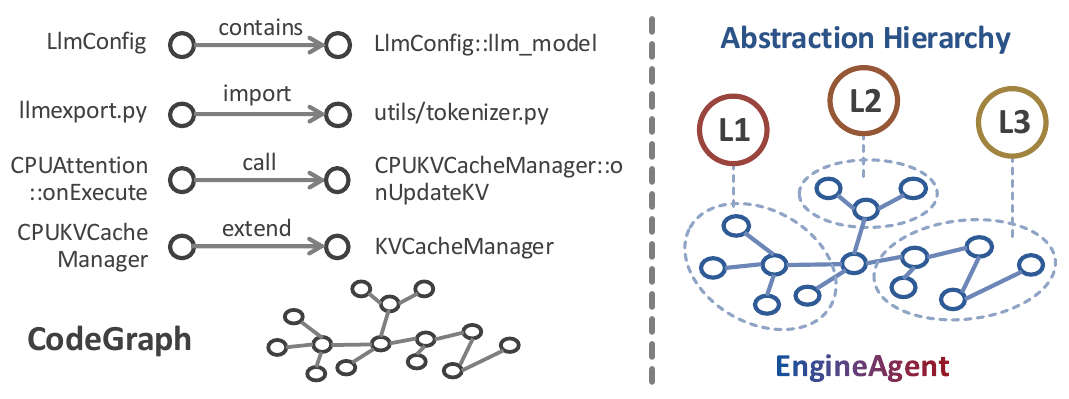}
    \caption{CodeGraph~\cite{CodeGraph} only captures syntactic structures rather than hierarchical understandin gs. Extracted relations in MNN CodeGraph as an example are illustrated on the left.}
    \label{fig:compare-graphs}
\end{figure}

\label{subsec:hierarchy-aware-design}
For a fixed engine snapshot, prior work CodeGraph~\cite{CodeGraph} provides a language-agnostic canonical graph of files, symbols, and typed static relations, including imports, calls, references, and instantiations, as illustrated in Figure \ref{fig:compare-graphs}. 
However, it captures only how code entities are structurally connected, without revealing the abstract knowledge of engine hierarchical design.

To encounter this, \systemname defines a nine-layer hierarchy based on the design abstraction of on-device LLM engines and automatically constructs a \emph{hierarchy-aware index} by semantically lifting the  CodeGraph representation of the engine into our pre-defined hierarchy.
In this way, engine abstractions are explicitly represented as structured semantic subgraph views over the original CodeGraph, extending syntactic relations to system-level understanding.

\paragraph{Hierarchy Abstraction}
As discussed in Section \ref{subsec:mobile-engines}, on-device LLM engines follow a two-stage workflow, where each stage contains three-level: input level, graph-level, and operator-level.
In order to better help agents understand the engine hierarchy, we further divide the three levels into nine layers, including \ding{172} chat template and tokenization and \ding{173} multimodal preprocessing layers on input level; \ding{174} model configuration, \ding{175} I/O tensor contracts, \ding{176} module mapping, \ding{177} weight sharing and KV sharing layers on graph level; and \ding{178} attention and KV cache, \ding{179} operator wrapping, and \ding{180} operator kernel layers on operator level.
Specifically, the offline-export has eight of these layers (L1--L8), spanning the export or compilation of each layer.
The execution-stage, on the other hand, contains all nine abstraction layers (L1--L9), extending the export-stage hierarchy with an additional L9 layer for hardware-specific kernels, covering the parsing and execution of exported artifacts on each layer.
For each abstraction layer, engine may have one or more instantiations: model-related layers (L3/L4) distinguish text-backbone and multimodal instantiations, while backend-dependent layers (offline L7--L8 and runtime L7--L9) are instantiated separately for CPU, Vulkan, OpenCL, and QNN when applicable.
We define these \emph{intra-layer} substructures as layer instantiations.
By defining this layered hierarchy in prompt, \systemname agentically construct a knowledge index from the raw CodeGraph to incorporate these knowledge into graph structures.


\paragraph{Agentic Index Construction and Review.}
\systemname constructs the hierarchy index as a nested graph structure following the predefined abstraction hierarchy.
The construction agent first divides the raw CodeGraph into export and execution stage subgraphs and then sequentially builds abstraction-layer subgraphs within each stage.
For each layer, the agent builds semantic understanding from the predefined hierarchy knowledge and repository evidence, selects representative code entities as seeds, and initializes the subgraph.
The subgraph is then expanded from these seeds by traversing CodeGraph relations and incorporating entities supported by additional evidence.
To improve reliability, \systemname separates semantic hierarchy attribution from raw graph construction, where static relation errors are checked independently while layer membership is determined through semantic evidence.
After construction, the indexing agent performs self-review and a read-only auditor validates the completeness and grounding of the generated subgraphs.
The resulting index provides multiple semantic views of the code entity graph across stages, abstraction layers, and instantiations.
By mapping engine functional abstractions onto concrete code entities, it bridges the gap between repository structure and engine organization, enabling agents to understand how high-level functionalities are realized by specific modules.

\paragraph{Example and Analysis (MNN Hierarchy Indexing)}

\begin{figure}[t]
    \centering
    \includegraphics[width=0.95\linewidth]{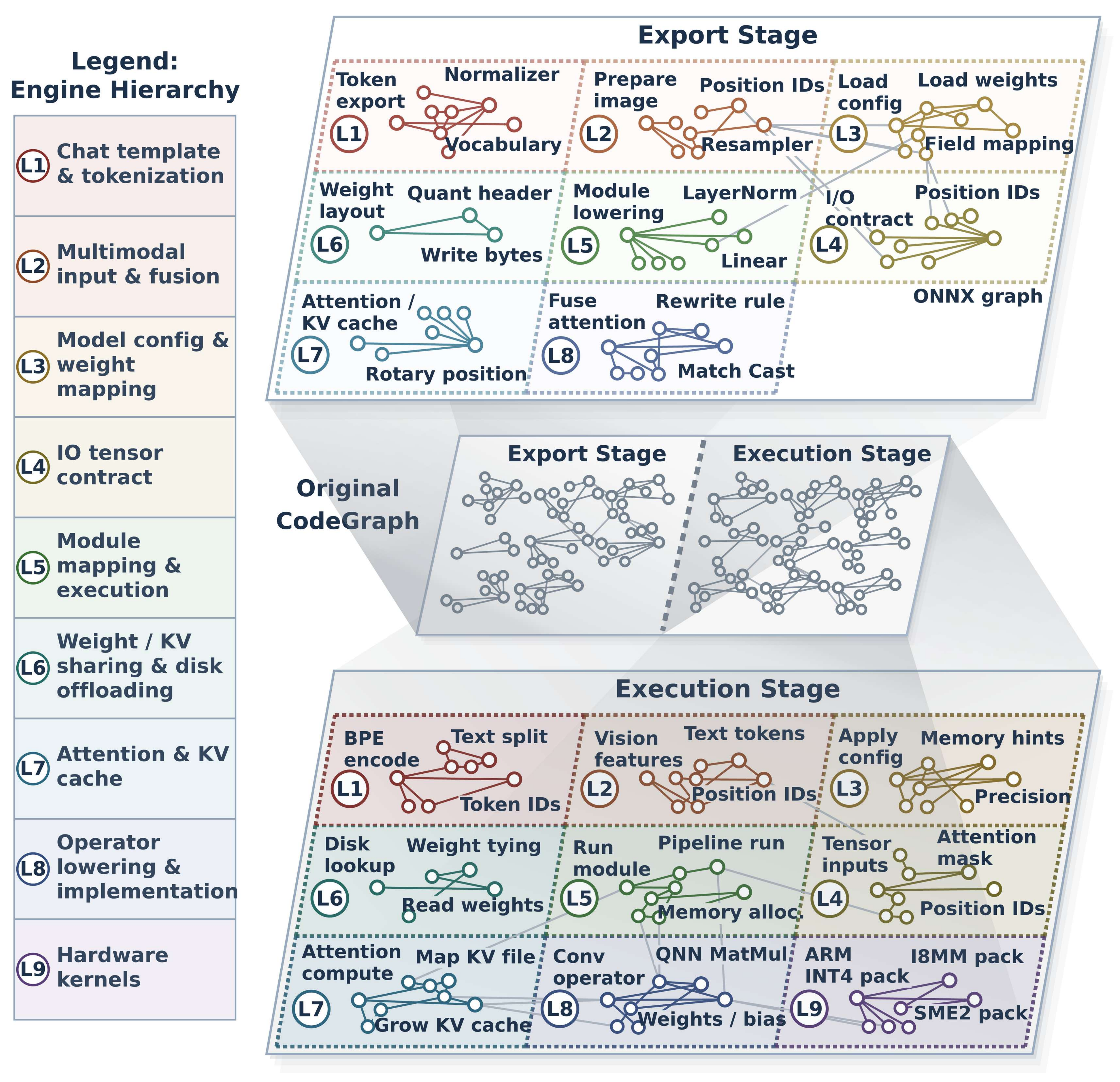}
    \caption{Hierarchy-aware engine modeling for MNN. The original CodeGraph (center) is organized into abstraction-layer subgraphs for export (top, L1--L8) and execution (bottom, L1--L9), with layer definitions shown on the left. Colored subgraphs group code entities by engine responsibility, while gray edges indicate cross-layer dependencies.}
    \label{fig:hierarchy-aware-graph}
\end{figure}

Figure~\ref{fig:hierarchy-aware-graph} visualizes the hierarchy-aware graph index constructed for MNN.
The middle panel shows the original CodeGraph, where files and symbols are connected only through static code relations.
After semantic lifting, \systemname reorganizes these entities into export- and execution-stage subgraphs and further into abstraction-layer subgraphs.
The resulting layers exhibit clear semantic coherence by grouping code entities according to their engine responsibilities.

\begin{figure}[h]
    \centering
    \includegraphics[width=0.82\linewidth]{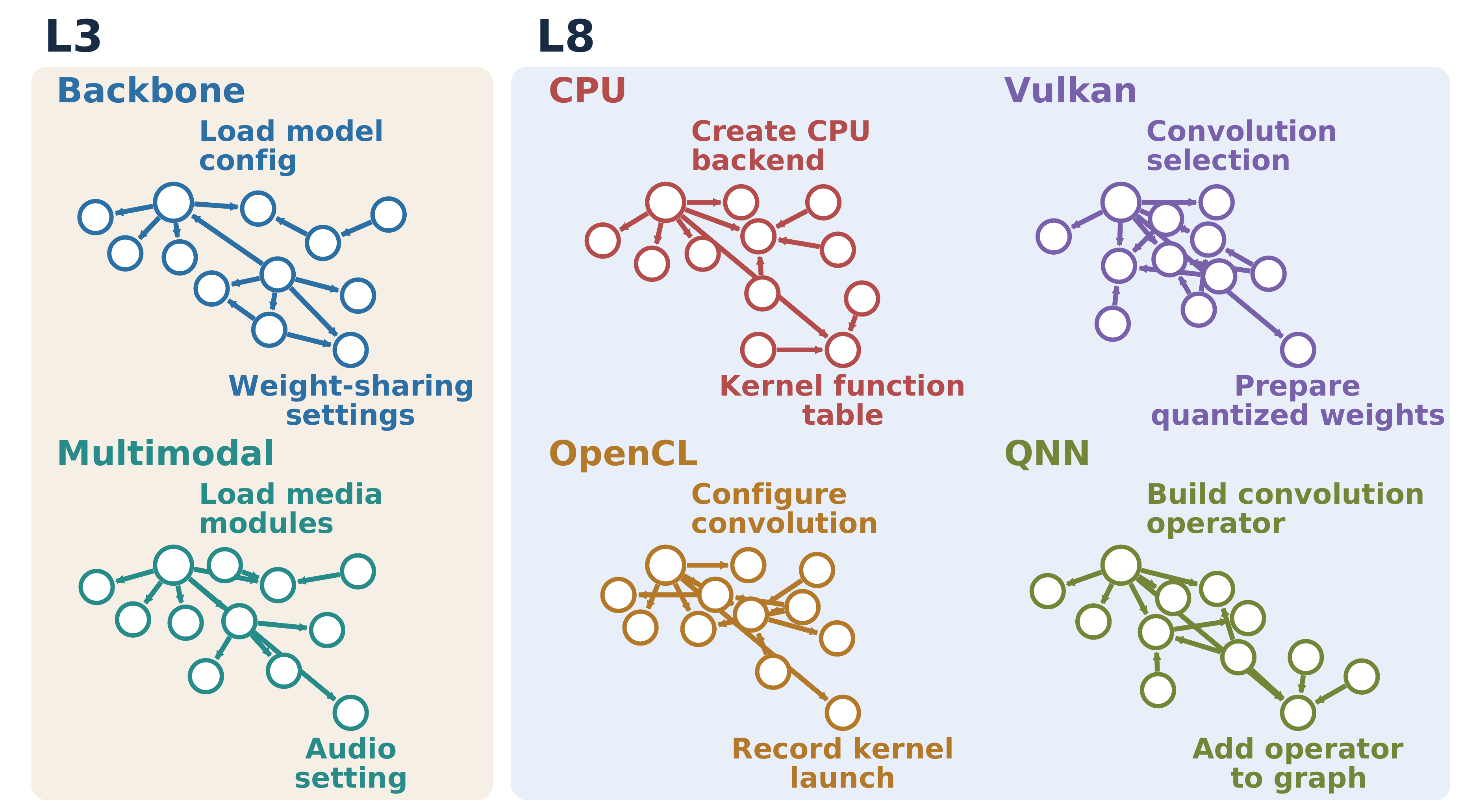}
    \caption{Instantiation subgraphs within MNN's execution-stage hierarchy, exposing modality- and backend-specific code structures within each layer.}
    \label{fig:mnn-instantiations}
\end{figure}


For the export stage, L1 groups tokenization-related entities such as \emph{Tokenizer export}, \emph{Normalizer}, and \emph{Vocabulary}; 
L2 captures multimodal preprocessing components such as \emph{Resampler}; 
L3 organizes model initialization logic including \emph{Model config} and \emph{Weight loading}; 
L5 groups module-level transformations such as \emph{LayerNorm} and \emph{Linear}; 
L7 captures attention-related components including \emph{Attention}, \emph{KV cache}, and \emph{Rotary position}; 
and L8 connects high-level operators with their lowered implementations, such as mapping Attention to \emph{Fuse attention}.
The execution-stage hierarchy further exposes corresponding runtime responsibilities:
L1 contains runtime token processing entities such as \emph{BPE encoder} and special token mapping;
L2 includes vision feature processing and multimodal fusion;
L5 groups runtime execution logic including module forward, graph traversal, and memory management;
L7 captures fused attention kernels and native flexible KV cache implementations;
L8 contains concrete operators such as \emph{Convolution} and \emph{QNN MatMul};
and L9 reaches hardware-specific implementations such as \emph{ARM INT4 pack}, \emph{I8MM}, and \emph{SME2 GEMM}. These examples demonstrate that the generated hierarchy is not a syntactic partition of the repository, but a semantic mapping between engine functionalities and concrete code entities.
It enables agents to reason about how high-level model functionalities are implemented across different abstraction layers.

Abstraction-layer subgraphs with multiple concrete realizations are further refined into subordinate \emph{instantiation subgraphs}.
As shown in Figure~\ref{fig:mnn-instantiations}, MNN's L3 separates backbone and multimodal implementations, while L8 branches into CPU, Vulkan, OpenCL, and QNN implementations with distinct backend-specific code paths.
These instantiation subgraphs expose how the same engine abstraction is realized differently across modalities and hardware backends.

\paragraph{Update Index under Engine Modification}
For each new commit, \systemname maps added, removed, and modified code entities onto the existing hierarchy-aware graph index and identifies the affected stage, abstraction-layer, and instantiation subgraphs.
Only these subgraphs and their immediate dependency neighborhoods are re-examined to update entity attribution, execution paths, and cross-layer relations, while unaffected subgraphs are directly reused. 
Only when a commit introduces major structural changes, or repeated incremental updates cause excessive accumulated drift, \systemname triggers a  hierarchy-aware index reconstruction.
Thus, incremental updates of the pre-constructed model-agnostic index incur only modest overhead.

\subsection{Host--Device Debugging and Verification Skills}
\label{subsec:hardware-aware-design}
Model enablement requires not only modifying the correct engine layers, but also repeatedly building, deploying, executing, and diagnosing the resulting implementation on real devices.
To support reliable hardware-aware verify-debug loop, \systemname agentically constructs dedicated verify-debug skills around these host--device workflows.
These procedures are largely engine and backend specific, yet remain reusable across subsequent new model enablements.

\paragraph{Automatic Skill Construction}
To provide agents with procedural knowledge for backend-specific building, verification, and failure diagnosis, \systemname agentically constructs \emph{build skills}, \emph{unit tests}, and \emph{few-shot debug skills} before individual model-enablement tasks.

\begin{figure}[!t]
    \centering
    \includegraphics[width=\linewidth]{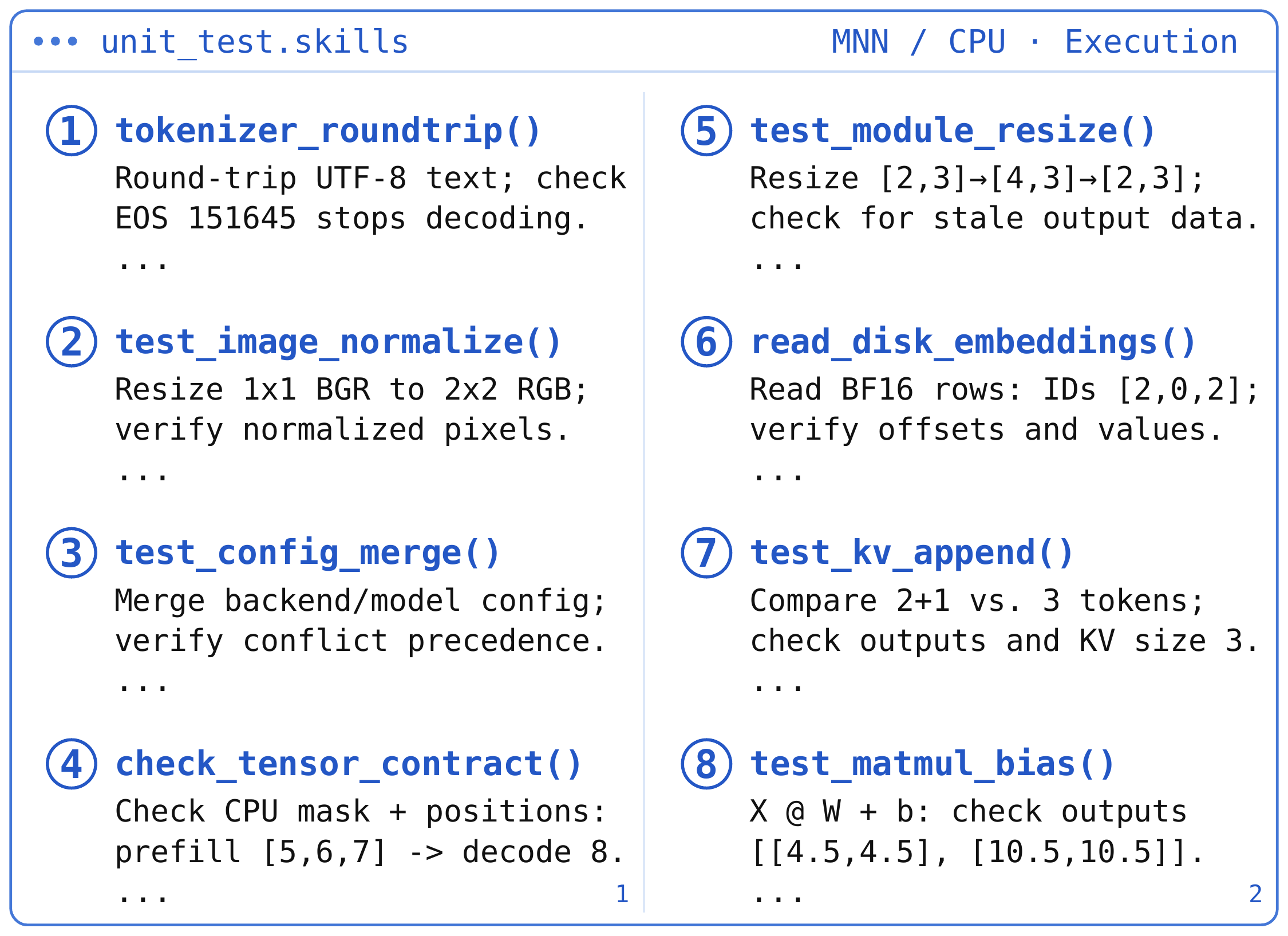}
    \caption{Examples of generated unit tests and debug skills for MNN-CPU execution stage by Deepseek-V4-Flash.}
    \label{fig:verify-debug}
\end{figure}

\emph{Build skills} organize information scattered across documentation, build scripts, and repository instructions into two executable workflows: model conversion/export and execution runner build and deployment.
Rather than merely summarizing documentation, the construction agent explores the relevant scripts and commands and validates the resulting workflow by exporting and executing an already supported model on the real device.

Besides, \emph{unit tests and few-shot debug skills} are  further constructed following the layer and instantiation division defined in hierarchy-aware graph index. 
For the export stage, it traverses L1--L8 in order and builds tests around the observable input/output contracts of each abstraction layer, from preprocessing and model configuration to module mapping and operator lowering.
For the execution stage, it similarly traverses L1--L9, extending validation through runtime operators and hardware-specific kernels.
When a layer contains multiple instantiations, tests are constructed separately for the corresponding modality or backend branches.
Figure \ref{fig:verify-debug} illustrate the example of generated skills for MNN-CPU execution stage.
These generated skills check and verify the outputs and functionality of each layer's modules. 
Together, these tests form a progressive validation chain, allowing failures to be localized to the earliest layer whose contract is violated rather than only being observed at end-to-end execution.
During this construction process, successful diagnostic procedures, characteristic failure symptoms, and corresponding recovery actions are summarized into \emph{few-shot debug skills} associated with their abstraction layers and hardware backends.

\paragraph{Update Skills under Engine Modification}
Prebuilt hardware aware verify-debug skills are updated using the same incremental principle as engine knowledge index updates.
Changes to build scripts, documentation, backend implementations, or toolchain configurations invalidate only their associated build skills, unit tests, and few-shot debug skills.
These affected artifacts are re-examined and revalidated, including replaying changed build or deployment workflows with an already supported model when necessary, while retaining unaffected artifacts.
A complete reconstruction is performed only when the corresponding backend workflow or toolchain undergoes substantial changes.

\subsection{Model-Specific Multi-Agent Main Loop}
\label{subsec:main-loop}
When a new model arrives, \systemname uses the prebuilt hierarchy index and debug skills to modify the target engine for a specified XPU backend.
This process is illustrated in Figure~\ref{fig:pipeline}.
\systemname instantiates three model-specific agents from the pre-constructed artifacts: a \emph{builder agent} for implementation, a \emph{debugger agent} for debugging reinforcement, and an independent \emph{auditor agent} for red-team verification.

The \emph{builder agent} first performs a \emph{difference analysis} against the engine's supported model set constructed during pre-enablement to identify the most similar reference models.
The architectural differences between the reference and new models are mapped onto the hierarchy-aware graph index to derive layer-wise modification plans.
The builder then implements these tasks sequentially across abstraction layers.
For each layer, the builder retrieves relevant hierarchy knowledge and hardware-aware \emph{build skills, unit tests, and few-shot debug skills} to guide implementation and self-validation.

The \emph{debugger agent} reinforces the builder when failures occur or progress becomes blocked.
Given the current abstraction layer, backend, execution logs, and validation results, it retrieves corresponding debugging skills, reproduces relevant workflows when necessary, and provides targeted diagnostic evidence and recovery suggestions.
By leveraging pre-constructed debug skills, the debugger avoids repeatedly exploring unstable verification paths during enablement.

The \emph{auditor agent} independently evaluates the builder's implementation as a red-team verifier.
After each abstraction layer is completed, it executes the corresponding pre-constructed unit tests and checks whether the implementation satisfies layer-specific requirements.
If validation fails, the auditor returns the earliest failing layer and device-side evidence for subsequent debugging.
After all layers pass, the auditor performs end-to-end validation covering model export, deployment, and on-device execution before approving the implementation.

Together, these agents form an iterative implementation--verification loop.
The builder proposes modifications, the auditor challenges their correctness, and the debugger provides targeted recovery when failures occur.
The loop continues until all layer-wise checks and end-to-end device validation pass.
Finally, the auditor-approved implementation is submitted to a rule-based non-agentic system verifier, which compares output logits against host-side PyTorch references and rejects implementations whose deviations exceed a predefined numerical tolerance.

\section{Evaluation}
\label{sec:evaluation}

\subsection{Benchmark Design}
\label{subsec:bench-design}
To evaluate the effectiveness of agentic systems for on-device LLM model enablement, we construct a benchmark comprising 12 tasks across four representative engine--backend pairs and three recent mobile LLMs: Qwen3.5-2B~\cite{Qwen3.5}, Gemma4-E2B-it~\cite{Gemma4}, and MiniCPM-V 4.6~\cite{MiniCPM-V-4.6}.
Each task requires enabling one target model on one engine--backend pair, covering diverse model architectures and heterogeneous CPU, GPU, and NPU execution paths.

\paragraph{Engine--Backend Pairs}
The benchmark covers four representative mobile engine--backend pairs:
\begin{itemize}
    \item MNN + CPU,
    \item llama.cpp + Vulkan,
    \item MLC-LLM + OpenCL,
    \item ExecuTorch + QNN.
\end{itemize}
These pairs span substantially different execution stacks and hardware abstractions.
MNN + CPU~\cite{Walle} represents a native CPU execution path with direct operator and kernel implementations; llama.cpp + Vulkan~\cite{llama.cpp} represents explicit GPU execution through Vulkan kernels and backend dispatch; MLC-LLM + OpenCL~\cite{MLC-LLM} captures a compiler-based deployment workflow targeting mobile GPUs; and ExecuTorch + QNN~\cite{ExecuTorch} represents graph lowering and execution through a vendor-specific NPU backend.
Together, they cover CPU, GPU, and NPU execution as well as native, compiler-based, and accelerator-specific enablement workflows.

\paragraph{Benchmark Case Construction and Engine Snapshot Selections}
For each model--engine--backend pair, if the target model has already been supported, we reset the engine to the commit immediately preceding the support commit and ask the agent to reproduce the enablement from this pre-support state.
Under this setting, the agent is not allowed to access future commits or search the web for post-snapshot implementations.
If the target model is not yet supported by the corresponding engine--backend pair, we use the latest available engine snapshot and construct a \emph{forward-enable} task.
Such tasks reflect realistic maintenance scenarios where no ground-truth support patch exists and success can only be determined through end-to-end validation.
Each benchmark task provides the agent with an engine repository snapshot, a target model checkpoint, the backend configuration, and validation scripts.
This setup requires the agent to solve the actual model-enablement problem rather than recover an existing implementation.

\paragraph{Success Evaluation and Criteria}
A task is considered successful only if the upgraded engine passes host compilation, device execution, and output-logit alignment checks verified by our evaluation program.
The evaluation program reproduces the provided build scripts to verify model export and runner cross-compilation, then deploys the generated artifacts to the target phone through ADB and checks that execution completes without errors.
It subsequently runs multiple on-device test cases covering all required modalities, collects the output logits, and compares them against host-side PyTorch reference results.
The implementation is accepted only when the logits agree with the reference outputs, with only negligible numerical drift attributed to hardware differences.

\subsection{Experimental Setup}
\label{subsec:eval-setup}

\paragraph{Baselines}
We compare against four representative coding-agent baselines.
\textbf{SWE-Agent}~\cite{SWE-Agent} represents repository-level autonomous software engineering, while \textbf{OpenCode}~\cite{OpenCode} and \textbf{Codex}~\cite{Codex} represent general-purpose coding agents.
We further construct a composite baseline, \textbf{OpenCode + CodeGraph + AgenTracer}, augmenting OpenCode with graph-based repository understanding~\cite{CodeGraph} and trajectory-level failure diagnosis~\cite{AgenTracer}.
Since coding agents frequently terminate prematurely on long-running enablement tasks without human continuation prompts, we enable the \emph{goal} mode of Codex and attach a lightweight continuous-run trigger to the other baselines, which resumes the agents when the evaluator has not yet accepted the implementation as successful and the time limit has not been reached.

\paragraph{Agent Model API}
SWE-Agent, OpenCode, and OpenCode + CodeGraph + AgenTracer use the DeepSeek-V4-Flash API~\cite{Deepseek-v4}, while Codex uses GPT-5.6 Sol~\cite{GPT-5}.
We evaluate \systemname with both DeepSeek-V4-Flash and GPT-5.6 Sol.
This provides matched comparisons against both groups of baselines and separates the effectiveness of the \systemname harness from LLM API it calls.

\paragraph{Metrics}
We primarily report \emph{end-to-end success rate}, together with five \emph{stage-wise success rates}: export, host text-only, host multimodal, device text-only, and device multimodal, to localize failures along the enablement pipeline. 
A stage is counted as successful only when all associated evaluator checks described above pass. We also report \emph{average wall-clock time} and \emph{average monetary cost} per task. Each run is capped at five hours, with timeouts counted as five hours; monetary cost is measured by the total model-API usage incurred during the run.

\subsection{Overall Performance across Benchmark Cases}
\label{subsec:e2e}
Table~\ref{tab:e2e-results} summarizes performance across 12 model-enablement tasks. \systemname achieves 9/12 end-to-end successes (75\%) with either DeepSeek-V4-Flash or GPT-5.6 Sol, substantially outperforming all baselines. The strongest baseline, Codex with GPT-5.6 Sol, succeeds on only 4/12 tasks (33.3\%), while OpenCode and CodeGraph + OpenCode + AgentTracer each succeed on 1/12 and SWE-Agent on none.

\begin{table}[h]
    \centering
    \caption{Overall model enablement performance across the 12 benchmark cases.}
    \label{tab:e2e-results}
    \resizebox{0.95\linewidth}{!}{
    \begin{tabular}{ccccc}
        \toprule
       \multirow{2}{*}{\makecell{\textbf{Agent}\\\textbf{Model API}}} &
        \multirow{2}{*}{\makecell{\textbf{Agentic}\\\textbf{Systems}}} &
        \multirow{2}{*}{\makecell{\textbf{E2E Success}\\\textbf{Rate}}} &
        \multicolumn{2}{c}{\textbf{Average}} \\
        \cmidrule(lr){4-5}
        &  &  &
        \textbf{Time} &
        \textbf{Cost (\$)} \\
        \midrule

        \multirow{5}{*}{\makecell{DeepSeek\\V4 Flash}}
        & SWE-Agent
        & 0/12
        & 4.85h
        & 2.11 \\

        & OpenCode
        & 1/12
        & 4.78h
        & 2.28 \\

        & \makecell{CodeGraph + OpenCode\\+ AgenTracer}
        & 1/12
        & 4.91h
        & 2.44 \\

        & \textbf{EngineAgent}
        & \textbf{9/12}
        & \textbf{3.82h}
        & \textbf{1.53} \\

        \midrule

        \multirow{2}{*}{GPT-5.6 Sol}
        & Codex
        & 4/12
        & 4.82h
        & 14.90 \\

        & \textbf{EngineAgent}
        & 9/12
        & 4.16h
        & 11.29 \\

        \bottomrule
    \end{tabular}
    }
\end{table}

\systemname achieves higher success rate while using less time and lower model-API cost than the baselines.
With DeepSeek-V4-Flash, \systemname completes each task in 3.82 hours on average at a cost of \$1.53, compared with 4.78--4.91 hours and \$2.11--\$2.44 for the corresponding baselines.
With GPT-5.6 Sol, \systemname averages 4.16 hours and \$11.29 per task, compared with 4.82 hours and \$14.90 for Codex.

\begin{table}[h]
    \centering
    \caption{Stage-wise model enablement success across the 12 benchmark cases. (T: Text-only backbone success rate on host/device, MM: Multi-Modal success rate on host/device)}
    \label{tab:stage-wise-results}
    \resizebox{\linewidth}{!}{
    \begin{tabular}{ccccccc}
        \toprule
        \multirow{3}{*}{\makecell{\textbf{Agent}\\\textbf{Model API}}} &
        \multirow{3}{*}{\makecell{\textbf{Agentic}\\\textbf{Systems}}} &
        \multicolumn{5}{c}{\textbf{Stage-wise Success Rate}} \\
        \cmidrule(lr){3-7}
        & &
        \multirow{2}{*}{\textbf{Export}} &
        \multicolumn{2}{c}{\textbf{Host}} &
        \multicolumn{2}{c}{\textbf{Device}} \\
        \cmidrule(lr){4-5}
        \cmidrule(lr){6-7}
        & & &
        \textbf{T} &
        \textbf{MM} &
        \textbf{T} &
        \textbf{MM} \\
        \midrule

        \multirow{5}{*}{\makecell{DeepSeek\\V4 Flash}}
        & SWE-Agent
        & 4/12 & 3/12 & 0/12 & 1/12 & 0/12 \\

        & OpenCode
        & 4/12 & 4/12 & 2/12 & 1/12 & 1/12 \\

        & \makecell{CodeGraph + OpenCode\\+ AgenTracer}
        & 4/12 & 4/12 & 4/12 & 1/12 & 1/12 \\

        & \textbf{EngineAgent}
        & \textbf{10/12} & \textbf{10/12} & \textbf{10/12}
        & \textbf{9/12} & \textbf{9/12} \\

        \midrule

        \multirow{2}{*}{GPT-5.6 Sol}
        & Codex
        & 6/12 & 6/12 & 4/12 & 6/12 & 4/12 \\

        & \textbf{EngineAgent}
        & \textbf{10/12} & \textbf{10/12} & \textbf{10/12}
        & \textbf{9/12} & \textbf{9/12} \\

        \bottomrule
    \end{tabular}
    }
\end{table}

Table~\ref{tab:stage-wise-results} shows where the performance gap emerges along the enablement pipeline. With DeepSeek-V4-Flash, \systemname achieves 10/12 successes in export and host-side text-only and multimodal execution, and 9/12 on both device-side paths. In contrast, DeepSeek-based baselines reach at most 4/12 at the host stage and fall to 0--1/12 on device. Adding CodeGraph and AgentTracer to OpenCode improves host-side multimodal success from 2/12 to 4/12, but device-side success remains 1/12, suggesting that stronger repository understanding alone is insufficient for hardware-grounded execution. With GPT-5.6 Sol, Codex reaches 6/12 on export and host text-only execution and 4/12 on multimodal execution, while \systemname still achieves 10/12 on host and 9/12 on device. Overall, \systemname's advantage extends beyond finding the correct implementation to maintaining correctness through multimodal integration and host-to-device execution.

\begin{figure}[h]
    \centering
    \includegraphics[width=0.75\linewidth]{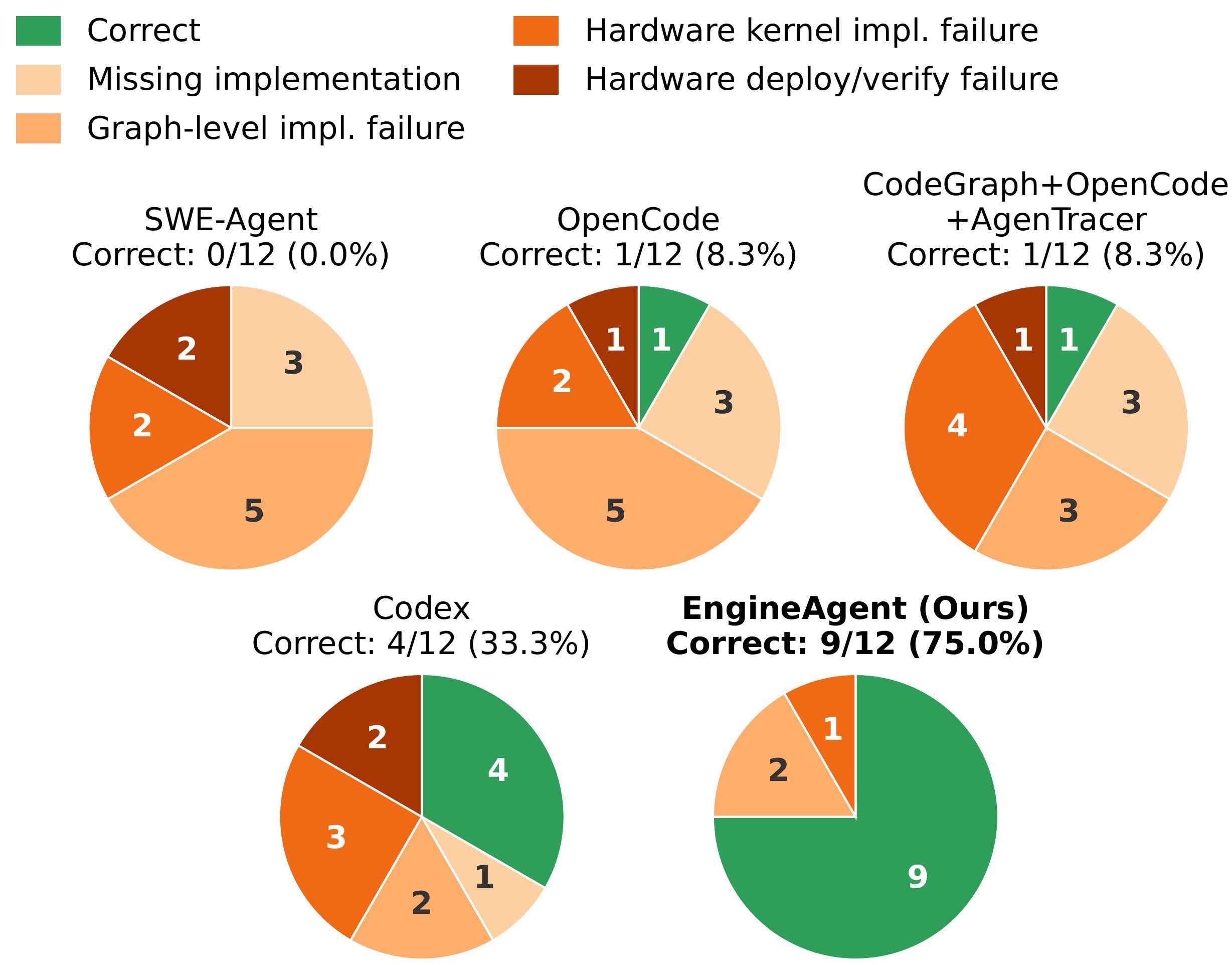}
    \caption{Failure breakdown across the 12 benchmark cases.}
    \label{fig:failure-analysis}
\end{figure}

Figure~\ref{fig:failure-analysis} breaks down failures across the 12 tasks. SWE-Agent and OpenCode each miss three required implementations and incur five graph-level failures, while Codex still fails in eight cases across all four categories. 
In contrast, \systemname eliminates both missing-implementation and hardware deployment/verification failures, and reduces hardware-kernel failures to 1/12. 
This suggests that hierarchy-aware knowledge helps recover the required implementation scope, while a stable build--verify--debug workflow prevents deployment failures and provides reliable feedback for hardware-specific debugging. 
As a result, \systemname's remaining three failures are limited to two graph-level and one hardware-kernel implementation failure.

\subsection{Breakdown Analysis By Cases}
\label{subsec:breakdown}

\begin{figure}[h]
    \centering
    \includegraphics[width=\linewidth]{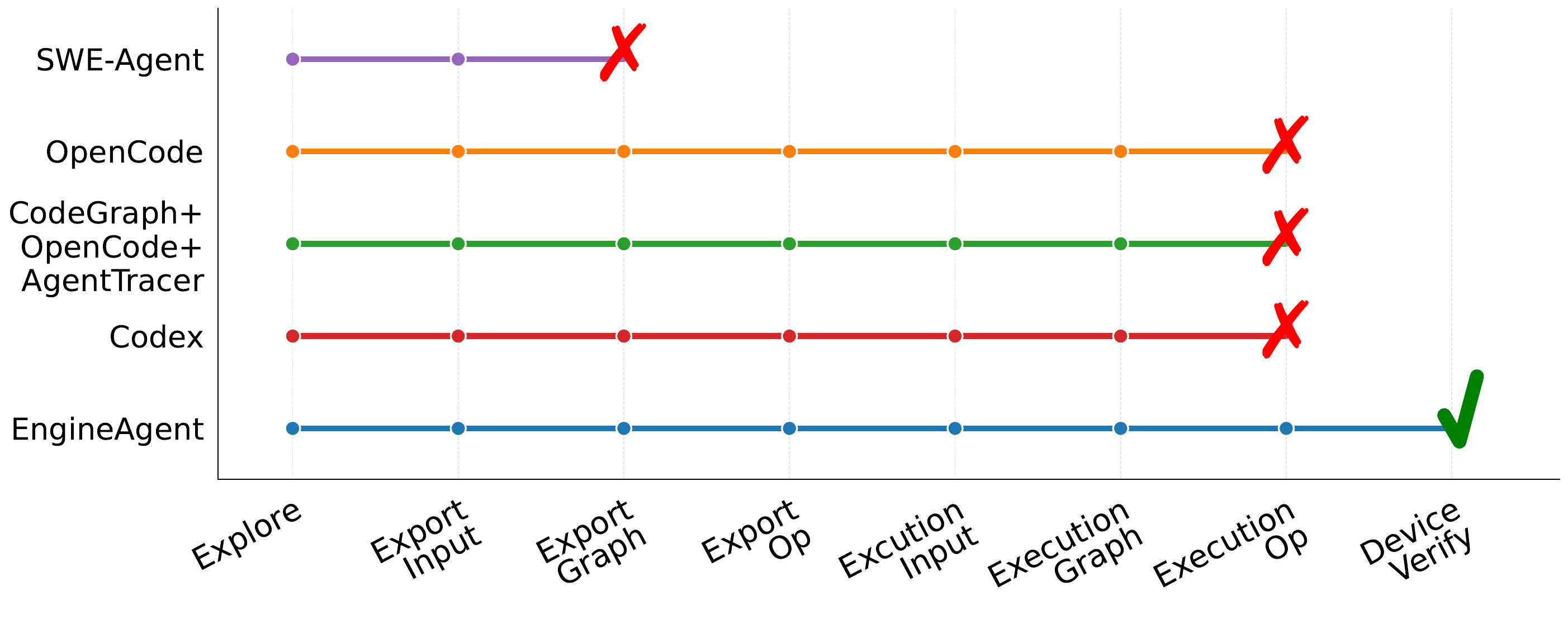}
    \caption{Progress Comparison of Agents enabling Qwen3.5-2B in llama.cpp-Vulkan.}
    \label{fig:progress_chart1}
\end{figure}

\paragraph{Case Study: Enabling Qwen3.5-2B on llama.cpp-Vulkan} 
The key challenge in enabling Qwen3.5-2B on llama.cpp-Vulkan is supporting the newly introduced Linear Attention operator on the Vulkan backend. As illustrated in Figure~\ref{fig:progress_chart1}, SWE-Agent fails at the graph-export stage because it overlooks an unsupported implementation required by the new operator. OpenCode, CodeGraph + OpenCode + AgentTracer, and Codex progress further but eventually fail when implementing the execution-stage Linear Attention kernel. In particular, these agents struggle to localize kernel-level implementation errors merely with their continuous end-to-end verification. 
In contrast, \systemname exploits its hardware-aware build-test-debug skills and directly constructs unit tests for the Linear Attention kernel and iteratively debugs the implementation on device. This hardware-grounded verification loop enables \systemname to progressively localize and fix the implementation errors, ultimately completing the enablement in approximately 4.5 hours.

\begin{figure}[h]
    \centering
    \includegraphics[width=\linewidth]{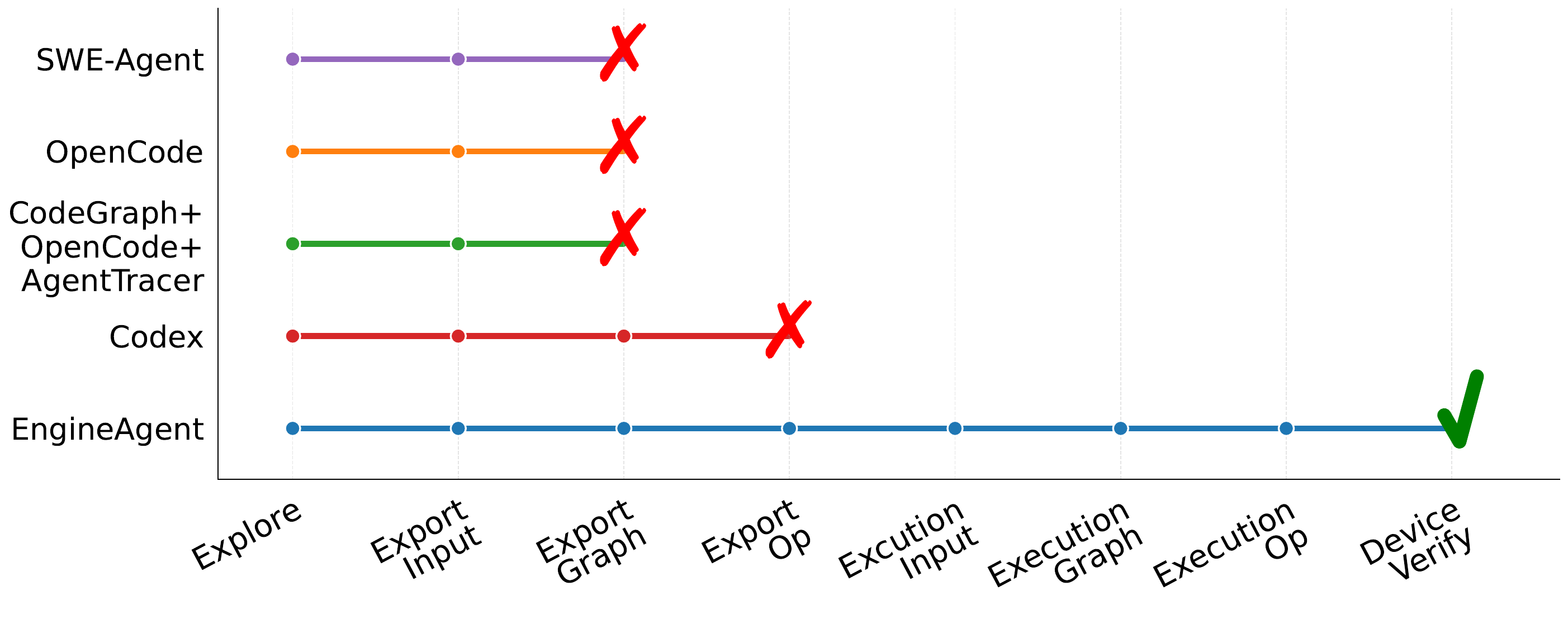}
    \caption{Progress Comparison of Agents enabling MiniCPM-V 4.6 in ExecuTorch-QNN.}
    \label{fig:progress_chart2}
\end{figure}

\paragraph{Case Study: Enabling MiniCPM-V 4.6 on ExecuTorch-QNN}
The key challenge in enabling MiniCPM-V 4.6 on ExecuTorch-QNN is to jointly modify the host-side offline QNN graph compilation and the online execution runtime. 
As illustrated in Figure~\ref{fig:progress_chart2}, all baselines fail during export: they repeatedly modify and debug the graph and operator export paths, but eventually time out because each iteration incurs the high latency of offline QNN preparation. 
In contrast, \systemname leverages its hierarchy-aware index to directly localize the required changes for the newly introduced window-attention \texttt{vit\_merger} in the Python exporter, successfully exporting the model within two attempts and completing the enablement at last.

\subsection{Cost Analysis for Pre-Constructed Index and Skills}
\label{subsec:construction-cost}
The engine hierarchy-aware index and hardware-aware skills are pre-constructed once and reused across model-enablement tasks.
As engines evolve, they are incrementally updated only the parts that are affected by each commit.
We evaluate both their initial construction and incremental update costs below.

\begin{table}[h]
    \centering
    \caption{Initial construction cost of pre-enablement artifacts on four engine snapshots.}
    \label{tab:construction-cost}
    \resizebox{0.75\linewidth}{!}{
    \begin{tabular}{cccc}
        \toprule
        \textbf{Engine} &
        \textbf{Snapshot} &
        \makecell{\textbf{Index Cost}\\\textbf{(\$)}} &
        \makecell{\textbf{Skills Cost}\\\textbf{(\$)}} \\
        \midrule
        MNN        & \texttt{f41d4638} & 1.04 & 1.83 \\
        llama.cpp  & \texttt{e06088da} & 0.66 & 2.27 \\
        MLC-LLM    & \texttt{84050697} & 0.77 & 2.40 \\
        ExecuTorch & \texttt{b7781552} & 1.25 & 2.68 \\
        \bottomrule
    \end{tabular}
    }
\end{table}

\paragraph{Initial Construction Costs.}
Table~\ref{tab:construction-cost} reports the initial construction costs for the four evaluated engines.
Constructing the hierarchy index costs \$0.66--\$1.25, while hardware-aware skills cost \$1.83--\$2.68, averaging \$0.93 and \$2.30, respectively.
Thus, the reusable pre-enablement artifacts require only modest one-time construction cost.

\begin{table}[h]
    \centering
    \caption{Update cost of pre-constructed artifacts across MNN versions.}
    \label{tab:mnn-update-cost}
    \resizebox{0.75\linewidth}{!}{
    \begin{tabular}{ccc}
        \toprule
        \textbf{Commit Update} &
        \makecell{\textbf{Index Update}\\\textbf{Cost (\$)}} &
        \makecell{\textbf{Skills Update}\\\textbf{Cost (\$)}} \\
        \midrule
        \texttt{f41d4638 $\rightarrow$ c5903082} & 0.0041 & 0.0000 \\
        \texttt{c5903082 $\rightarrow$ 0bff03cb} & 0.0195 & 0.1262 \\
        \texttt{0bff03cb $\rightarrow$ 17c00ba6} & 0.0051 & 0.0000 \\
        \texttt{17c00ba6 $\rightarrow$ 4419e33b} & 0.0047 & 0.0507 \\
        \texttt{4419e33b $\rightarrow$ 30dfb9bb} & 0.0049 & 0.1459 \\
        \midrule
        \textbf{Average} & \textbf{0.0077} & \textbf{0.0646} \\
        \bottomrule
    \end{tabular}
    }
\end{table}

\paragraph{Update Costs.}
Table~\ref{tab:mnn-update-cost} reports the incremental cost across five consecutive MNN updates.
On average, updating the hierarchy index costs only \$0.0077 and updating the hardware-aware skills costs \$0.0646 per commit, corresponding to only 0.7\% and 3.5\% of their initial construction costs, respectively.
Combined, each update costs only \$0.072 on average, about 2.5\% of the initial pre-construction cost, demonstrating that the reusable artifacts can be maintained with very low overhead as the engine evolves.

\subsection{Ablation Study}
\label{subsec:ablation}

\begin{figure}[h]
    \centering
    \subfigure[w/o hierarchy index]{
        \includegraphics[width=0.45\linewidth]{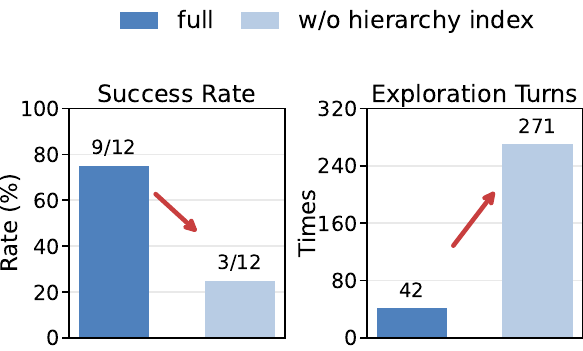}
        \label{fig:ablation-hierarchy}
    }
    \hfill
    \subfigure[w/o hardware skills]{
        \includegraphics[width=0.47\linewidth]{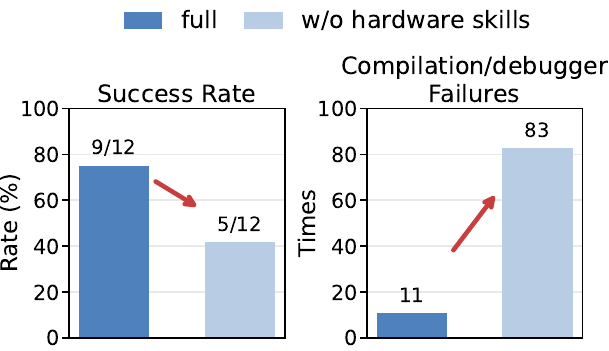}
        \label{fig:ablation-skills}
    }
    \caption{Ablation study of the hierarchy-aware index and hardware-aware skills.}
    \label{fig:ablation}
\end{figure}

We ablate the two pre-constructed components of \systemname to quantify their contributions. As shown in Figure~\ref{fig:ablation}, removing the \emph{hierarchy-aware index} reduces end-to-end success from 9/12 to 3/12 and increases average exploration turns from 42 to 271, showing that explicit hierarchy knowledge is critical for recovering complete implementation scopes and cross-layer dependencies. Removing the \emph{hardware-aware skills} reduces success from 9/12 to 5/12 and increases compilation/debugging failures from 11 to 83, highlighting the importance of pre-validated build, verification, and debugging workflows for reliable hardware-grounded execution. Together, these results show that the two components are complementary: hierarchy-aware indexing guides complete implementation, while hardware-aware skills enable reliable on-device validation and debugging.

\section{Related Works}
\label{sec:related-works}

\subsection{General-Purpose Coding Agents}
\label{subsec:auto-SDE}
Commercial coding agents~\cite{GitHubCopilot,Cursor,ClaudeCode,Codex}, SWE-agent~\cite{SWE-Agent}, and OpenHands~\cite{OpenHands} integrate repository exploration, editing, and execution feedback.
Agentless~\cite{Agentless} structures issue resolution into localization, repair, and validation.
Repository understanding draws on dependency-aware planning~\cite{CodePlan}, code graphs~\cite{RepoGraph,LocAgent,CodexGraph}, and program structure~\cite{AutoCodeRover}.
Other work supports interactive debugging~\cite{DebugGym}, reusable skills learned from execution traces~\cite{GSE}, and tree search with agent-based evaluation~\cite{SWESearch}.

\subsection{Mobile Programming and Deployment Agents}
AutoIOT~\cite{AutoIOT} and GPIoT~\cite{GPIoT} synthesize AIoT programs from natural-language requirements, while AgentDeploy~\cite{AgentDeploy} partitions models across existing inference engines and reuses optimization configurations for mobile deployment. At the operator level, MoKA, introduced in MobileKernelBench~\cite{MobileKernelBench}, combines repository-aware reasoning with iterative implementation and on-device validation. EngineAgent targets complete support for previously unsupported LLM and multimodal architectures, requiring coordinated changes across model conversion, graph construction, runtime-state management, and backend execution, validated through end-to-end execution on the target device.

\subsection{Automation of ML Systems}
\label{subsec:auto-MLsys}
TVM~\cite{TVM}, Ansor~\cite{Ansor}, Roller~\cite{Roller}, and Welder~\cite{Welder} automate tensor compilation and optimization through graph transformations, scheduling, search, and fusion.
KernelBench~\cite{KernelBench} evaluates LLM-generated GPU kernels, while GEAK~\cite{GEAK} and KernelAgent~\cite{KernelAgent} guide kernel optimization with execution feedback and hardware information.

\section{Conclusion}
\label{sec:conclusion}

Supporting rapidly evolving LLMs on heterogeneous mobile hardware makes model enablement a recurring cross-layer systems challenge. We presented \systemname, which separates reusable engine and hardware knowledge from model-specific implementation. Its hierarchy-aware index guides coordinated changes across engine abstractions, while hardware-aware workflows and a builder--auditor loop ground implementation and debugging in device-side validation. Our evaluation shows that this design improves model-enablement success while reducing time and model-API cost compared with coding-agent baselines. More broadly, \systemname offers a path to turn specialized integration expertise into reusable infrastructure, reducing the gap between model innovation and deployment across diverse hardware.




\bibliographystyle{ACM-Reference-Format}
\bibliography{ref-eurosys}

\clearpage

\nobalance

\end{document}